\documentclass[a4paper,11pt]{article}

\usepackage{contribution}
\usepackage{amsmath}
\usepackage{amsfonts}
\usepackage{amssymb}
\usepackage{epsfig}

\newcommand{\weblink}[2][]{%
    \ifthenelse{\equal{#1}{}}%
    {\textnormal{\url{#2}}}%
    {\textnormal{\href{#2}{#1}}}%
}

\newcommand{\acknowledgements}[1]{%
  \bigskip\bigskip
  \textsf{\textbf{\Large Acknowledgements}} \\[2ex]
  {#1}
  \bigskip
}

\def\beq{\begin{equation}}
\def\eeq#1{\label{#1}\end{equation}}
\def\eeqn{\end{equation}}

\def\beqa{\begin{eqnarray}}
\def\eeqa#1{\label{#1}\end{eqnarray}}
\def\eeqan{\end{eqnarray}}

\let\bar=\overbar

\def\Dslash{\not{\hbox{\kern-4pt $D$}}}
\def\dslash{\not{\hbox{\kern-2pt $\del$}}}

\def\ee{e^+e^-}

\def\msb{{\bar{\ssstyle M \kern -1pt S}}}

\def\beq{\vspace{-0.3em}\color{magenta}\begin{eqnarray*}}
\def\eeq{\end{eqnarray*}\color{blue}\vspace{-0.3em}}
\def\bea{\begin{eqnarray}}
\def\eea{\end{eqnarray}}
\def\be{\begin{equation}}
\def\ee{\end{equation}}

\def\bfsigma{\mbox{\boldmath $\sigma$}}

\def\lQ{\Lambda_{\rm QCD}}
\def\em{{\rm em}}

\def\als{\alpha_{\rm s}}

\def\siml{{\ \lower-1.2pt\vbox{\hbox{\rlap{$<$}\lower6pt\vbox{\hbox{$\sim$}}}}\ }}     
\def\simg{{\
    \lower-1.2pt\vbox{\hbox{\rlap{$>$}\lower6pt\vbox{\hbox{$\sim$}}}}\ }}

\newcommand{\contribution}[7][]{%
  \clearpage
  \thispagestyle{plain}
  \ifthenelse{\equal{#1}{}}
  {\hypersetup{pdftitle={#2}}}
  {\hypersetup{pdftitle={#1}}}
  \hypersetup{pdfauthor={{#3} {#4}}}
  {\centering\normalfont\LARGE\bfseries\sffamily #2 \par\nobreak}
  \lhead{}
  \chead{%
    \textit{\footnotesize XIV International Conference on Hadron Spectroscopy
      (\weblink[\textit{hadron2011}]{http://www.hadron2011.de}), 13-17 June 2011, Munich, Germany}%
  }
  \rhead{}
  \bigskip
  \begin{center}
    {#3} {#4}\ifthenelse{\equal{#6}{}}{}{\footnote{\weblink[#6]{mailto:#6}}}
    \ifthenelse{\equal{#7}{}}{}{#7} \\
    \textit{#5}
  \end{center}
  \bigskip
}

\renewcommand{\abstract}[1]{%
  \begin{center}
    \begin{minipage}{0.85\textwidth}
      \begin{footnotesize}
        #1
      \end{footnotesize}
    \end{minipage}
  \end{center}
  \bigskip
}

\begin{document}

\contribution[EFTs for Quarkonium and Dipole Transitions]  
{Effective Field Theories for Quarkonium \\ and Dipole Transitions}  
{Antonio}{Vairo}  
{Physik-Department\\
 Technische Universit\"at M\"unchen\\
 James-Franck-Str. 1, 85748 Garching, Germany}  
{antonio.vairo@ph.tum.de}  
{}

\abstract{
Effective field theories for quarkonium at zero and finite temperature provide an unifying description 
for a wide class of phenomena. As an example, we discuss physical effects induced by dipole transitions.
}

\section{Hierarchies}

\vspace*{-11.0cm}

\hfill TUM-EFT 23/11

\vspace*{10.5cm}

Quarkonia, i.e. heavy quark-antiquark bound states, are systems characterized by hierarchies of energy scales \cite{Brambilla:2004wf}.
They follow from the quark mass, $M$, being the largest scale in the system, which, in particular, means that $M \gg p$, the typical momentum 
transfer in the system, $M \gg \lQ$, the hadronic scale, and $M \gg \pi T \gg \hbox{other thermal scales}$, 
where $T$ is the temperature of the medium. 
These hierarchies allow systematic studies through the construction of suitable effective field theories (EFTs).

{\it (i) The non-relativistic expansion}\\
$M \gg p$ implies that quarkonia are non-relativistic and characterized by the hierarchy of scales typical of a non-relativistic bound state:
$p \sim 1/r \sim Mv$  and $E \sim M v^2$, where $r$ is the typical radius, $E$ the typical binding energy and 
$v \ll 1$ the heavy-quark velocity in the centre-of-mass frame.  Note that the hierarchy of non-relativistic scales makes the very
difference of quarkonia with heavy-light mesons, which are characterized just by the two scales $M$ and $\lQ$.

Systematic expansions in the small heavy-quark velocity $v$ may be implemented at the
Lagrangian level by constructing suitable non-relativistic effective field theories (EFTs) \cite{Brambilla:2004jw}.

{\it (ii) The perturbative expansion}\\
$M \gg \lQ$ implies $\als(M) \ll 1$: phenomena happening  at the scale $M$ may be treated perturbatively.
We may further have small couplings if $Mv \gg \lQ$ and $Mv^2 \gg \lQ$, 
in which case  $\als(Mv) \ll 1$ and $\als(Mv^2)\ll 1$ respectively. Moreover, we have $v \sim \als(Mv)$.
This is likely to happen only for the lowest charmonium and bottomonium states, which 
may be described by weakly-coupled Coulombic bound states, while excited quarkonia probe 
the transition from Coulombic to confined bound states.

{\it (iii) The thermal expansion}\\
If the temperature of the medium in heavy-ion collisions is such that $M \gg \pi T$, which is the 
case for most present days colliders, this implies that the quarkonium remains a non-relativistic 
bound state also in the thermal bath induced by the medium. 
However, the temperature will, in general, interfere with the other scales of the bound state. 
As a consequence, bound state observables like masses, lifetimes, decay widths etc. will be modified 
by the medium. In particular, it is expected that at sufficiently high temperatures the interference of the 
medium will be such to dissociate the quarkonium. Since different quarkonia have different radii and different 
binding energies, different quarkonia are expected to dissociate in the medium at different temperatures, 
providing a thermometer for the plasma \cite{Matsui:1986dk}, see also \cite{hadLai}.
$\pi T \gg \hbox{other thermal scales}$ implies a hierarchy also in the thermal scales.

\section{Effective field theories}
\begin{figure}[ht]
\makebox[0cm]{\phantom b}
\put(40,0){\epsfxsize=12truecm\epsffile{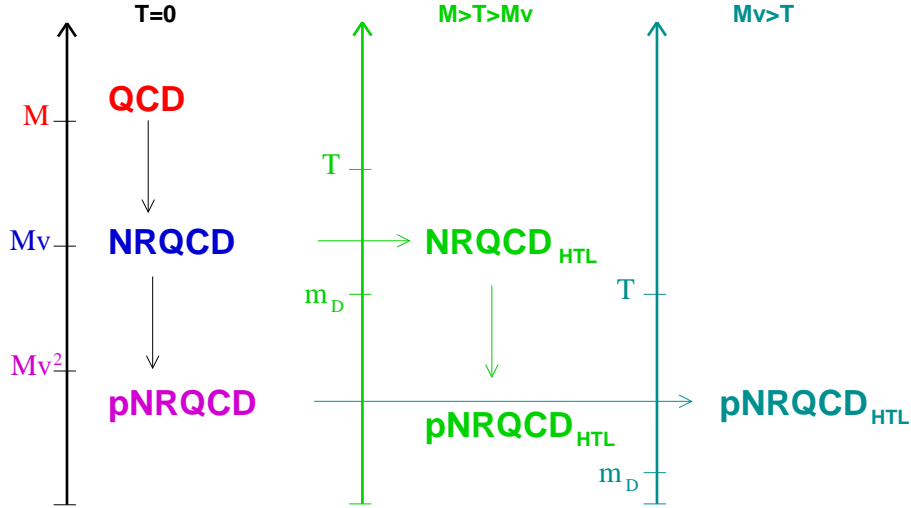}}
\caption{Hierarchies of EFTs for quarkonium at zero temperature \cite{Brambilla:2004jw} and at finite temperature 
\cite{Brambilla:2008cx,Escobedo:2008sy,Vairo:2009ih,Brambilla:2010vq,Escobedo:2010tu}.
}
\label{figeft}
\end{figure}

The hierarchies of EFTs for quarkonium at zero and finite temperature are shown in Fig.~\ref{figeft}.
In the following, we will consider systems for which $Mv \gg T$, so that both the scale $M$ and the scale $Mv$ may 
be integrated out ignoring  medium effects (third column of Fig.~\ref{figeft}).

Heavy quark-antiquark annihilation and production happen at the scale $M$. The suitable EFT is NRQCD \cite{Caswell:1985ui,Bodwin:1994jh}. 
The effective Lagrangian is organized as an expansion in $1/M$ and $\als(M)$: 
\be
{\cal L}_{\rm NRQCD}  = \sum_n \frac{c_n(\als(M),\mu)}{M^{n}} \times O_n(\mu)\,, 
\label{nrqcd}
\ee
where $O_n$ are NRQCD operators of dimension $4+n$ and $c_n$ are NRQCD matching coefficients. 
For quarkonium production in NRQCD, see also \cite{hadBut}.

The heavy quark and antiquark in quarkonium cannot be resolved at scales lower than $Mv$.
The suitable EFT is pNRQCD \cite{Pineda:1997bj,Brambilla:1999xf}. 
The effective Lagrangian is organized as an expansion in  $1/M$, $\als(M)$  and $r$:
\be
 {\cal L}_{\rm pNRQCD}  =  \int d^3r\,  
\sum_n \sum_k  \frac{c_n(\als(M),\mu)}{M^{n}}  \times 
 V_{n,k}(r,\mu^\prime, \mu) \; r^{k}  \times O_{n,k}(\mu^\prime) \,,
\label{pnrqcd}
\ee
where $O_{n,k}$ are pNRQCD operators and $V_{n,k}$ are the pNRQCD matching coefficients.
The matching coefficients of the four-fermion, dimension six, operators may be interpreted as the potentials 
of the bound-state Schr\"odinger equation, while the matching coefficients of the higher-dimension 
operators describe the couplings of the heavy quarks to the low-energy degrees of freedom.

\begin{figure}[ht]
\begin{center}
\epsfxsize=15truecm \epsffile{{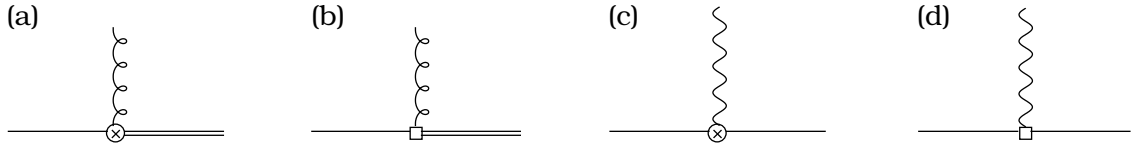}}
\caption{Chromoelectric (a), chromomagnetic (b), electric (c) and magnetic (d) dipole interaction vertices. The single line stands for a 
colour-singlet quark-antiquark propagator, while the double line for a colour-octet quark-antiquark propagator.}
\label{figdip}
\end{center}
\end{figure} 

To list the low-energy degrees of freedom and to write explicitly the Lagrangian of pNRQCD we need to specify 
our system. In the following, we will concentrate on the physics of the quarkonium ground states in the presence 
of a medium whose temperature is much lower than the typical moment transfer in the bound state (this situation includes the va\-cuum).
For a recent review, also on the physics of the quarkonium ground states, we refer to \cite{Brambilla:2010cs}.
The suitable EFT for the quarkonium ground states is weakly coupled pNRQCD, since for those systems 
$Mv \sim M\als \gg Mv^2 \sim M\als^2 \simg \lQ$.
The degrees of freedom are quark-antiquark states (colour singlet, S, colour octet, O), 
low-energy gluons and photons, and $n_f$ light quarks ($q_i$). The Lagrangian reads
\bea
{\cal L}_{\rm pNRQCD}   &=& \int d^3r \; {\rm Tr}\, \left\{ {\rm S}^\dagger \left( i\partial_0 -  \frac{{\bf p}^2}{M} + \dots -  V_s \right){\rm S} 
+  {\rm O}^\dagger \left( i{D_0} - \frac{{\bf p}^2}{M} + \dots -  V_o \right){\rm O}\right\}
\nonumber\\
&& 
- \frac{1}{4}   F_{\mu\nu}^a F^{\mu\nu\,a} - \frac{1}{4}   F_{\mu\nu} F^{\mu\nu} 
+ \sum_{i=1}^{n_f}\bar{q}_i\,iD\!\!\!\!/\,q_i  + \Delta {\cal L}\,.
\eea

At leading order in the power counting, the singlet field S satisfies a Schr\"odinger equation with potential $V_s$.
Higher-order terms are in $\Delta {\cal L}$, which describes the interaction with the low-energy degrees of freedom.
The leading interactions are (chromo)electric and (chromo)magnetic dipole interactions
($ee_Q$ is the electric charge of the heavy flavour $Q$):
\bea
\Delta {\cal L} &=&
\int d^3r  \;  {\rm Tr} \Bigg\{  
V_A{\rm O^\dagger}{ \bf r}\cdot g {\bf E}\,{\rm S}
+ \cdots
+ \frac{1}{2 M}\;V_1\; \left\{{\rm S}^\dagger , \bfsigma \cdot g {\bf B}\right\} {\rm O}
+ \cdots
\nonumber\\
&& +   V_A^\em \; {\rm S}^\dagger { \bf r}\cdot e e_Q {\bf E}^{\em} {\rm S}   
+ \cdots
+ \frac{1}{2  M}\;V_1^\em\; \left\{{\rm S}^\dagger , \bfsigma \cdot e e_Q {\bf B}^{\em}\right\} {\rm S}
+ \dots \Bigg\}\,.
\label{dipole}
\eea
The corresponding Feynman diagram vertices are shown in Fig.~\ref{figdip}. The matching coefficients 
$V_A$, $V_1$, $V_A^\em$ and $V_1^\em$ are one at leading order in the coupling.

In the following, we will consider the effect of the self-energy correction to the singlet propagator induced 
by the dipole vertices \eqref{dipole} in three different observables: the quark-antiquark static energy at zero temperature
in perturbation theory, the photon line shape in the $J/\psi \to X\,\gamma$ radiative decay for 0 MeV $\le E_\gamma \siml$ 500 MeV and the 
$\Upsilon(1S)$ width induced by a medium whose temperature is about twice the critical temperature.

\begin{figure}[ht]
\begin{center}
\epsfxsize=14truecm \epsffile{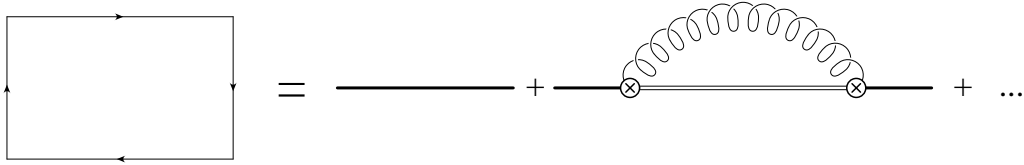}
\caption{The Wilson loop in the large time limit (left side) in terms of the pNRQCD singlet propagator (right side).}
\label{figwil}
\end{center}
\end{figure} 

\section{The perturbative potential and static energy at $T=0$}
The quark-antiquark static energy, $E_0$, is given by the large-time exponential fall off of the 
static Wilson loop \cite{Susskind:1976pi}. In pNRQCD, the large-time Wilson loop is matched 
by the singlet propagator, see Fig.~\ref{figwil}. Hence, the static energy is given 
by the singlet static potential $V_s^{(0)}$ plus corrections due to the coupling of the singlet 
to low-energy gluons and light quarks. The one-loop correction is shown in the right side of Fig.~\ref{figwil}:
the low-energy gluon is coupled to the singlet through the chromoelectric dipole vertex of Fig.\ref{figdip}(a).
Explicitly the static energy is given by 
\bea
&& E_0(r) = 
V_{s}^{(0)}(r, \mu)  - 
\hbox{$\displaystyle 
i \frac{g^2}{3} V_A^2  
\int_0^\infty dt \, e^{-i t  (V_o^{(0)}-V_s^{(0)})} \,   
\langle  {\rm Tr}\{{\bf r}  \cdot {\bf E}(t) \, {\bf r}  \cdot  {\bf E}(0)\} \rangle(\mu)  $} + \dots \,,
\label{static}
\eea
where the chromoelectric correlator $\langle  {\rm Tr}\{{\bf r}  \cdot {\bf E}(t) \, {\bf r}  \cdot  {\bf E}(0)\} \rangle$
comes from the two chromoelectric dipole vertices. 
The factorization scale, $\mu$, dependence cancels between the two terms in the right-hand side, 
therefore, the $\mu$ dependence of the singlet static potential, 
$V_s^{(0)} \sim \ln r\mu, \ln^2 r\mu$, ..., may be deduced from the 
$\mu$ dependence of the one loop correction in pNRQCD $ \sim \ln (V_o^{(0)}-V_s^{(0)})/\mu, \ln^2 (V_o^{(0)}-V_s^{(0)})/\mu, ... 
  \ln r\mu, \ln^2 r\mu, ...\,$.

Since the static Wilson loop is known up to N$^3$LO \cite{Schroder:1998vy,Brambilla:1999qa,Anzai:2009tm,Smirnov:2009fh}, 
the octet potential, $V_o^{(0)}$, is known up to NNLO \cite{Kniehl:2004rk,Brambilla:2010xn}, 
$V_A = 1 + {\cal O}(\als^2)$ \cite{Brambilla:2006wp} and 
the chromoelectric correlator  $\langle  {\rm Tr}\{{\bf r}  \cdot {\bf E}(t) \, {\bf r}  \cdot  {\bf E}(0)\} \rangle$
is known up to NLO \cite{Eidemuller:1997bb}, 
from \eqref{static} it follows that up to N$^4$LO (in the scheme of \cite{Brambilla:2006wp}) 
\bea
 V_s^{(0)}(r,\mu)   &=& 
- \frac{4}{3}\frac{\als(1/r)}{r} 
\left[1 + {\tilde a}_1  \frac{\als(1/r)}{4\pi} + {\tilde a}_{2\, s} \left(\frac{\als(1/r)}{4\pi}\right)^2 
\right.
\nonumber\\
&& \hspace{-12mm}
+ \left(144\,\pi^2 \,  \ln {r\mu}   + {\tilde a}_{3\, s}\right)\, \left(\frac{\als(1/r)}{4\pi}\right)^3 
\nonumber\\
&& \hspace{-12mm}
\left.
+ \left(  a_{4}^{L2} \ln^2 {r \mu} 
         +  \left(a_{4}^{L}  
         +  48 \pi^2 \,\beta_0 (- 5 + 6 \ln 2)\right)  \ln {r\mu}  + {\tilde a}_{4\, s} \right)
\left(\frac{\als(1/r)}{4\pi}\right)^4 
\right]\,,
\eea
where the coefficient ${\tilde a}_1$ may be read from \cite{Fischler:1977yf,Billoire:1979ih}, 
${\tilde a}_{2\, s}$ from \cite{Schroder:1998vy}, ${\tilde a}_{3\, s}$ from \cite{Anzai:2009tm,Smirnov:2009fh}, 
$a_{4}^{L2}$ and  $a_{4}^{L}$ from \cite{Brambilla:2006wp}, while ${\tilde a}_{4\, s}$ is unknown. 
The potentially large logarithms, $\ln {r\mu}$, may be resummed by solving the corresponding renormalization group 
equations; the static potential at N$^3$LL then reads \cite{Pineda:2000gza,Brambilla:2009bi}:
\bea
V_{s}^{(0)}(r,\mu)  &=&  V_{s}^{(0)}(r,1/r)  + \frac{8}{9}  r^2 \left[ V_o^{(0)}(r,1/r)  -  V_s^{(0)}(r,1/r)\right]^3
\nonumber\\
&& \hspace{3cm}
\times \left( \frac{2}{\beta_0}  \ln \frac{ \als(\mu)}{\als(1/r)} 
+ \eta_0\left[ \als(\mu)  -  \als(1/r)  \right] \right)\,, 
\\
\eta_0 &\equiv& \frac{1}{\pi}\left[-\frac{\beta_1}{2\beta_0^2} + \frac{12}{\beta_0}
\left(\frac{-5n_f + 18 \pi^2+ 141}{108} \right)\right] \,,
\eea
where $\beta_i$ are the coefficients of the beta function.

\begin{figure}[h]
\begin{center}
\epsfxsize=9truecm \epsfbox{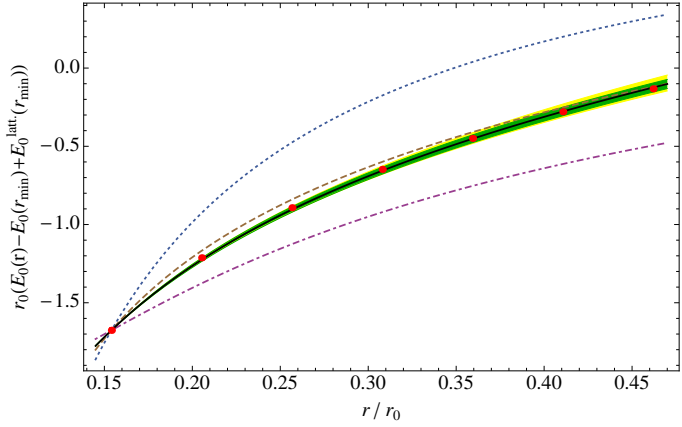}
\caption{The static quark-antiquark energy at N$^3$LL taken from \cite{Brambilla:2010pp} plotted against 
the quenched lattice data from \cite{Necco:2001xg}. $r_0$ stands for a lattice scale of dimension $-1$.}
\label{figlat}
\end{center}
\end{figure}

Finally, summing back the low-energy contributions in \eqref{static}, we obtain 
the static quark-antiquark energy at N$^3$LL \cite{Brambilla:2009bi}, which  
may be compared with lattice data (see Fig.~\ref{figlat}).
The conclusion is that perturbation theory, supplemented by a suitable renormalon subtraction scheme,  
describes well the static quark-antiquark energy at short distances, i.e. up to distances of about 0.25 fm 
($r_0 \approx 0.5$ fm in physical units). Indeed, one can use this to extract $\Lambda^{n_f=0}_{\rm \overline{MS}}r_0=0.622^{+0.019}_{-0.015}$ 
and, in perspective, $r_0$, once high-precision unquenched lattice data will be available \cite{Brambilla:2010pp}.

\section{The photon line shape in $J/\psi \to X\,\gamma$ for 0 MeV $\le E_\gamma \siml$ 500 MeV}
We consider the radiative decay $J/\psi \to X\,\gamma$ for 0 MeV $\le E_\gamma \siml$ 500 MeV.
The relevant scales are: $p \sim 1/r \sim M_cv \sim$ 700 MeV - 1 GeV $> \lQ$, 
$E_{J/\psi}\equiv  M_{J/\psi}-2M_c  \sim M_cv^2 \sim$ 400 MeV - 600 MeV and  
0 MeV $\le E_\gamma \siml$ 500 MeV, which is smaller than $M_cv$. 
It follows that the system is { \it (i)} non-relativistic, 
{ \it (ii)} weakly-coupled at the scale $M_cv$: $v \sim \als$, and 
{ \it (iii)} that we may multipole expand in the external photon energy \cite{Brambilla:2005zw}. 

Three main processes contribute to $J/\psi \to X\,\gamma$ for 0 MeV $\le E_\gamma \siml$ 500 MeV.

\begin{figure}[htb]
\begin{center}
\epsfxsize=7truecm \epsfbox{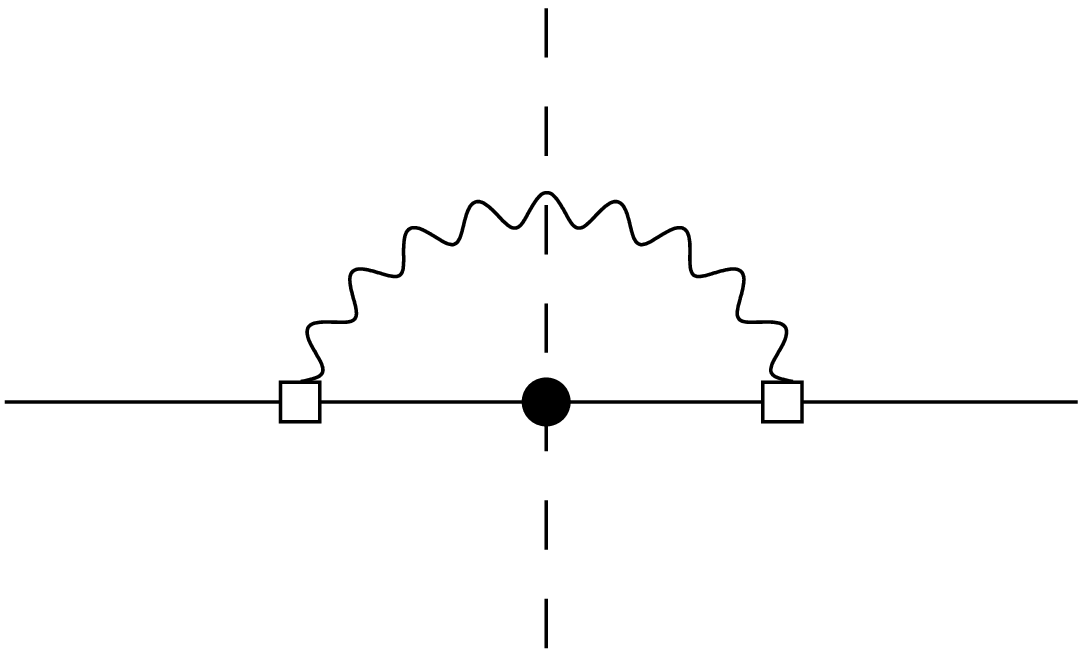}
\caption{Magnetic dipole transition induced by the vertex of Fig.~\ref{figdip}(d). The black dot stands for the 
imaginary part of $V_s$, which is responsible for the decay of the $\eta_c$.}
\label{figM1}
\end{center}
\end{figure}

{\it (i) Magnetic dipole transition $J/\psi \to \eta_c \,\gamma \to X\,\gamma$}\\
The $J/\psi$ may decay through an intermediate magnetic dipole transition to an $\eta_c$ and a photon.
This process is shown by the cut diagram in Fig.~\ref{figM1}. The differential width reads
\be 
\frac{d\Gamma_{\rm mag}}{dE_\gamma} = \frac{64}{27}
\frac{\alpha}{\pi}\frac{E_\gamma}{M_{J/\psi}^2} \frac{\Gamma_{\eta_c}}{2} 
\frac{E_\gamma^2}{(M_{J/\psi}-M_{\eta_c}-E_\gamma)^2 + \Gamma_{\eta_c}^2 /4}\,.
\label{Gammam}
\ee
$\Gamma_{\eta_c} \sim M_{c}\als^5$ is the $\eta_c$ width; for $\Gamma_{\eta_c}\to 0$ one recovers 
$\displaystyle \Gamma(J/\psi \to \eta_c \,\gamma) = \frac{64}{27} \alpha  \frac{E_\gamma^3}{M_{J/\psi}^2}$.
We observe that the non-relativistic Breit--Wigner distribution goes like:
\be
\frac{E_\gamma^2}{(M_{J/\psi}-M_{\eta_c}-E_\gamma)^2 + \Gamma_{\eta_c}^2 /4}
= \left\{ 
\begin{array}{cc}
1 & \hbox{for} ~~ E_\gamma \gg M_c\als^4 \sim M_{J/\psi}-M_{\eta_c} \\
\frac{E_\gamma^2}{(M_{J/\psi}-M_{\eta_c})^2} &  \hbox{for} ~~ 
E_\gamma \ll M_c\als^4 \sim M_{J/\psi}-M_{\eta_c} 
\end{array}
\right..
\ee

\begin{figure}[htb]
\begin{center}
\epsfxsize=7truecm \epsfbox{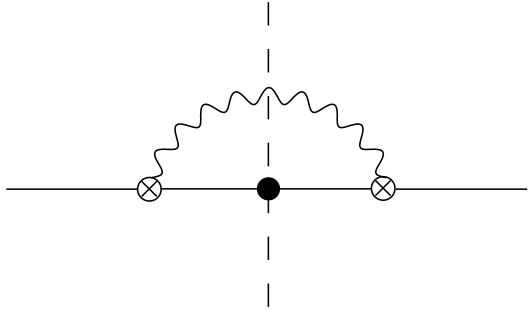}
\caption{Electric dipole transition induced by the vertex of Fig.~\ref{figdip}(c). The black dot stands for the 
imaginary part of $V_s$, which is responsible for the decay of the $\chi_{c0,2}(1P)$.}
\label{figE1}
\end{center}
\end{figure}

{\it (ii) Electric dipole transition $J/\psi \to \chi_{c0,2}(1P) \,\gamma \to X\,\gamma$}\\
The $J/\psi$ may decay through an intermediate electric dipole transition to a $\chi_{c0,2}$ and a photon.
This process is shown by the cut diagram in Fig.~\ref{figE1}. The differential width reads \cite{Voloshin:2003hh}
\bea
\frac{d\Gamma_{\rm ele}}{dE_\gamma} &=& 
\frac{7168}{6561}
\frac{\alpha}{\pi}\frac{E_\gamma}{M_{J/\psi}} \als^5 \, |a(E_\gamma)|^2\,,
\label{Gammae}
\\
a(E_\gamma) &\equiv& \frac{(1-\nu)(3+5\nu)}{3(1+\nu)^2} 
\nonumber\\
&& +\frac{8\nu^2(1-\nu)}{3(2-\nu)(1+\nu)^3}\,{}_2F_1(2-\nu,1;3-\nu;-(1-\nu)/(1+\nu))\,,
\nonumber\\
\nu  &\equiv& \sqrt{-E_{J/\psi}/(E_\gamma-E_{J/\psi})}\,.
\nonumber
\eea
Since 
\be
|a(E_\gamma)|^2
= \left\{ 
\begin{array}{cc}
1 & \hbox{for} ~~ E_\gamma \gg M_c\als^2 \sim E_{J/\psi} \\
{E_\gamma^2}/{(2E_{J/\psi})^2} &  \hbox{for} ~~ 
E_\gamma \ll M_c\als^2 \sim E_{J/\psi} 
\end{array}
\right.\,,
\ee
${d\Gamma_{\rm mag}}/{dE_\gamma}$ and ${d\Gamma_{\rm ele}}/{dE_\gamma}$ are of equal order for 
$M_c\als \gg E_\gamma \gg M_c \als^2 \sim -E_{J/\psi}$; 
the magnetic contribution dominates for 
$ -E_{J/\psi} \sim M_c \als^2 \gg E_\gamma \gg M_c\als^4 \sim  M_{J/\psi}-M_{\eta_c}$;
it also dominates by a factor 
$E^2_{J/\psi}/( M_{J/\psi}-M_{\eta_c})^2 \sim 1/\als^4$ for $E_\gamma \ll  M_c\als^4 \sim M_{J/\psi}-M_{\eta_c}$.
In practice, since $|a(E_{J/\psi})|^2 \approx 0.075$,  the magnetic dipole transition 
$J/\psi \to \eta_c \,\gamma \to X\,\gamma$ is the dominant process over the whole range 
0 MeV $\le E_\gamma \siml$ 500 MeV.

{\it (iii) Fragmentation}\\
Fragmentation and other background processes are typically modeled and fitted to the data.\\

\begin{figure}[ht]
\centering
\includegraphics[scale=0.32, angle=-90]{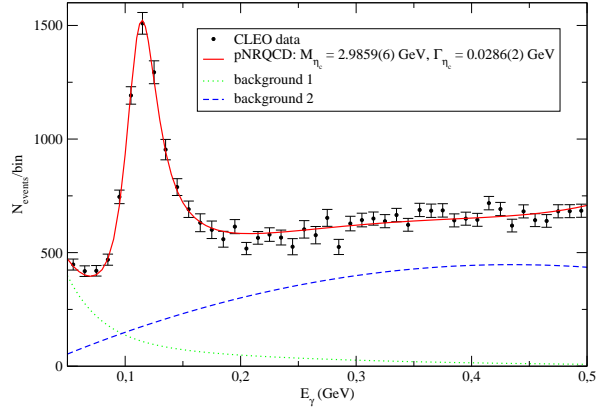}
\caption{Fit of $d(\Gamma_{\rm mag}+\Gamma_{\rm ele})/dE_\gamma$  
 plus background \cite{Brambilla:2010ey} on the CLEO data of \cite{Mitchell:2008fb}.}
\label{figfit}
\end{figure}

Fitting \eqref{Gammam} plus \eqref{Gammae} plus background 
on the CLEO data of \cite{Mitchell:2008fb}, we get Fig.~\ref{figfit} \cite{Brambilla:2010ey}. 
The line-shape parameters are 
\be
M_{\eta_c} =  2985.9 \pm 0.6\,\hbox{(fit)} \, \hbox{MeV} \,, 
\qquad\qquad \Gamma_{\eta_c} = 28.6\pm 0.2\,\hbox{(fit)}   \, \hbox{MeV}\,,
\ee
where theoretical errors have not been included.
Besides $M_{\eta_c}$ and $\Gamma_{\eta_c}$ the fitting
parameters are the overall normalization, the signal normalization, 
and (three) background parameters.

A study of electric transition in quarkonium in pNRQCD has been presented in \cite{hadPie}.

\section{$\Upsilon(1S)$ thermal width for $T \siml 2T_c$}
The bottomonium vector ground state, $\Upsilon(1S)$, produced in heavy-ion collisions at the LHC 
may possibly realize the hierarchy \cite{Vairo:2010bm} (see also \cite{hadGhi})
$$
M_b \approx 5 \; \hbox{GeV} \; 
> M_b \als \approx 1.5 \; \hbox{GeV} \; >  \pi T \approx 1 \; \hbox{GeV} \; 
>  M_b \als^2 \approx 0.5  \; \hbox{GeV} \; 
\simg  m_D, \lQ \,,
$$
where $T$ is the temperature of the QCD plasma created by the collisions.
A temperature $T$, such that $\pi T$ is of the order of 1 GeV, is about twice the critical temperature of the quark-gluon 
plasma formation, $T_c$; $m_D$ stands for the next-relevant thermal scale: the Debye mass.
Studies of the $\Upsilon(1S)$ properties and, in particular, of its width in the above conditions are very timely
because signals of bottomonium dissociation have just been seen by the CMS experiment \cite{Silvestre:2011ei}.

According to the above hierarchy, the bound state is weakly coupled, 
the temperature is lower than $M_b \als$, implying that the bound state is mainly Coulombic, 
and the effects due to the scale $\lQ$ and to the other thermodynamical scales may be neglected.

\begin{figure}[ht]
\begin{center}
\epsfxsize=8truecm \epsfbox{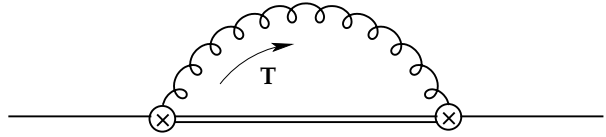}
\caption{Leading thermal contribution to the singlet propagator from the scale $T$.}
\label{figT}
\end{center}
\end{figure}

Integrating out $T$ from pNRQCD modifies pNRQCD into pNRQCD$_{\rm HTL}$ (see Fig.~\ref{figeft}), whose
Yang--Mills Lagrangian gets an additional hard thermal loop (HTL) part \cite{Braaten:1991gm} 
and potentials get additional thermal corrections. One effect of the HTL part is to give a 
mass, $m_D$, to the temporal gluons. The leading thermal contribution to the potential 
is encoded in the diagram of Fig.~\ref{figT}, where thermal gluons couple to the singlet 
through chromoelectric dipole vertices (the difference with the diagram in Fig.~\ref{figwil} is in the 
gluon propagator). The loop momentum region is taken to be $k_0\sim T$  and $k \sim T$.

\begin{figure}[ht]
\begin{center}
\epsfxsize=8truecm \epsfbox{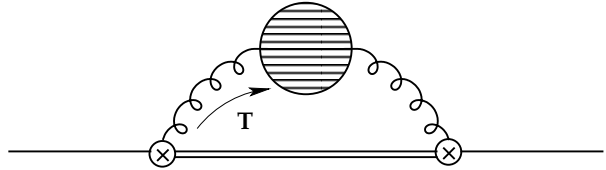}
\caption{Gluon self-energy correction to the diagram of Fig.~\ref{figT}.}
\label{figself}
\end{center}
\end{figure}

The gluon self-energy correction to the diagram in Fig.~\ref{figT} is shown 
in Fig.~\ref{figself}. This diagram has an imaginary part that contributes to the 
thermal width of the state:
\bea
\Gamma_{1S}^{(T)}  &=&  \left[ - \frac{4}{3} \als T m_D^2
\left( -\frac{2}{\epsilon} + \gamma_E + \ln\pi 
- \ln\frac{T^2}{\mu^2}+ \frac{2}{3} - 4 \ln 2 - 2 \frac{\zeta^\prime(2)}{\zeta(2)} \right) \right.
\nonumber\\
&& \left.
\hspace{7.3cm} -\frac{32\pi}{3}  \,  \als^2\, T^3 \, \ln 2\right] a_0^2\,,
\label{GammaT}
\eea
where $\displaystyle a_0 = \frac{3}{2M_b \als}$.
The width is infrared (IR) divergent; the divergence has been regularized in dimensional regularization
($D=4+\epsilon$).

\begin{figure}[ht]
\begin{center}
\epsfxsize=11truecm \epsfbox{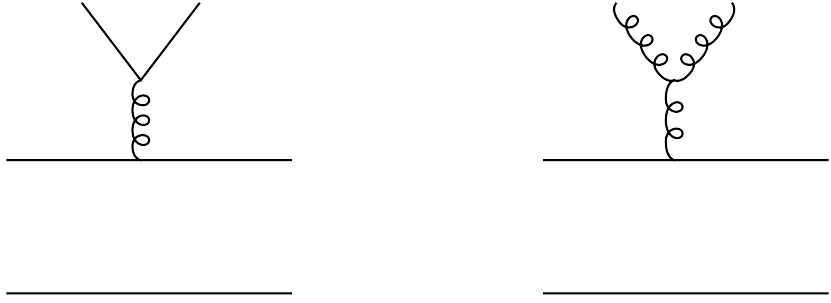}
\caption{Landau-damping scatterings.}
\label{figlandau}
\end{center}
\end{figure}

The origin of this thermal width may be traced back to the Landau-damping phenomenon, i.e. the scattering of 
heavy quarks with hard space-like particles in the medium (see Fig.~\ref{figlandau}).
The Landau-damping phenomenon plays a crucial role in quarkonium dissociation \cite{Laine:2006ns}.
It is when ${\rm Im}~ V_s(r)\vert_{\rm Landau-damping} \sim {\rm Re}~ V_s(r) \sim \als/r$ that the quarkonium dissociates.
The dissociation temperature is parametrically given by $\pi T_{\rm dissociation} \sim M_b  g^{4/3}$.
Note that the interaction is screened when $1/r  \sim m_D$ and that in the weak coupling 
($m_D \sim gT$) $\pi T_{\rm screening} \sim M_b  g \gg \pi T_{\rm dissociation}$.
The typical dissociation temperature, $T_{\rm dissociation}$, for the  $\Upsilon(1S)$
is about 450 MeV \cite{Escobedo:2010tu}, which implies that a temperature, $T$, such that 
$\pi T$ is about 1 GeV, is below the dissociation temperature.

\begin{figure}[ht]
\begin{center}
\epsfxsize=8truecm \epsfbox{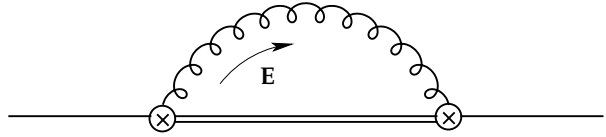}
\caption{Leading thermal contribution to the singlet propagator from the scale $E$. Gluons are HTL gluons.}
\label{figE}
\end{center}
\end{figure}

Integrating out the energy scale $E$ from pNRQCD$_{\rm HTL}$ 
provides corrections to the mass and width of the quarkonium in the thermal bath.
The leading diagram is shown in Fig.~\ref{figE}, where HTL gluons couple to the singlet 
through chromoelectric dipole vertices. The loop momentum region is taken to be  
$k_0\sim E$  and $k \sim E$. 
For $E \gg m_D,\lQ$, the contribution to the thermal width of the $\Upsilon(1S)$ is given by 
\bea
\Gamma_{1S}^{(E)}  &=&
4\als^3T-\frac{64}{9M_b}\als TE_1+\frac{32}{3}\als^2T\frac{1}{M_b a_0} +\frac{7225}{162} E_1\als^3
\nonumber\\
&&
-\frac{4 \als Tm_D^2}{3}\left(\frac{2}{ \epsilon}
+\ln\frac{E_1^2}{\mu^2}+\gamma_E-\frac{11}{3}-\ln\pi+\ln4\right) a_0^2 
+\frac{128\als Tm_D^2}{81}\frac{\als^2}{E_1^2}\,I_{1,0}\,,
\label{GammaE}
\eea
where $\displaystyle E_1=-\frac{4 M_b \als^2}{9}$ and $I_{1,0}=-0.49673$ (similar to the Bethe logarithm).
The width is ultraviolet (UV) divergent. Note that the 
UV divergence  of \eqref{GammaE} cancels against the IR
divergence of \eqref{GammaT}.

\begin{figure}[ht]
\begin{center}
\epsfxsize=5truecm \epsfbox{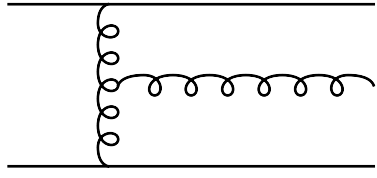}
\caption{Singlet-to-octet break up diagram.}
\label{figso}
\end{center}
\end{figure}

The thermal width $\Gamma_{1S}^{(E)}$, which is of order $\als^3 T$, 
is generated by the break up of a quark-antiquark colour-singlet state 
into an unbound quark-antiquark colour-octet state (see e.g. Fig.~\ref{figso}):
a process that is kinematically allowed only in a medium.
The singlet to octet break up is, therefore, a different phenomenon with respect 
to the Landau damping. In the situation $M_b \als^2 \gg m_D$, the first dominates over the second 
by a factor $(M_b \als^2/m_D)^2$ \cite{Brambilla:2008cx}.

The complete thermal width up to ${\cal O}(m\als^5)$ is \cite{Brambilla:2010vq}:
\bea
\Gamma_{1S}^{(\mathrm{thermal})}  &=& \Gamma_{1S}^{(T)}+\Gamma_{1S}^{(E)} =
\frac{1156}{81}\als^3T+\frac{7225}{162}E_1\als^3+\frac{32}{9} \als\, Tm_D^2\,a_0^2\,I_{1,0}
\nonumber\\
&&
-\left[\frac{4}{3} \als T m_D^2
\left(\ln\frac{E_1^2}{T^2}+ 2\gamma_E -3 -\ln 4- 2 \frac{\zeta^\prime(2)}{\zeta(2)} \right)
+\frac{32\pi}{3} \,  \als^2\, T^3 \, \ln 2 \right] a_0^2\,.
\eea
The width is an observable, therefore, finite and scheme independent. The logarithm, 
$\ln E_1^2/T^2$, is a relic of the cancellation between the IR divergence at the scale $T$ 
and the UV divergence at the scale $E$.

\section{Conclusions}
Our understanding of the theory of quarkonium has dramatically improved 
over the last fifteen years. An unified picture has emerged that is able to describe large 
classes of observables for quarkonium in the vacuum and in a medium.
For the ground state, precision physics is possible and lattice data 
provide often a crucial complement. In the case of quarkonium in a hot medium,
systematic treatments have disclosed new phenomena that may eventually be responsible 
for the quarkonium suppression observed in heavy-ion collisions.

\acknowledgements{
I acknowledge financial support 
from the DFG cluster of excellence ``Origin and structure of the universe'' 
(\href{http://www.universe-cluster.de}{www.universe-cluster.de}) and 
from the DFG project BR4058/1-1 
``Effective field theories for strong interactions with heavy quarks''.
}

\end{document}